\def\PR{{Phys. Rev.\ }\/}
\def\PRL{{ Phys. Rev.Lett.\ }\/}

\def\JMMM{{ J. Magn. Magn. Mater.}\ }
\def\etal{{\it et.al.}\/}

\def\be{\begin {equation}}
\def\ee{\end {equation}}
\def\ber{\begin {eqnarray}}
\def\eer{\end {eqnarray}}
\def\bers{\begin {eqnarray*}}
\def\eers{\end {eqnarray*}}

\makeatletter

\newcommand{\Rmnum}[1]{\expandafter\@slowromancap\romannumeral #1@}
\makeatother

\makeatletter
\newcommand*\env@matrix[1][*\c@MaxMatrixCols c]{%
  \hskip -\arraycolsep
  \let\@ifnextchar\new@ifnextchar
  \array{#1}}
\makeatother

\documentclass[aps,prb,showpacs,superscriptaddress,a4paper,twocolumn]{revtex4-1}

\usepackage{amsmath}
\usepackage{float}
\usepackage{color}

\usepackage{amsfonts}
\usepackage{amssymb}
\usepackage{graphicx}
\begin {document}

\title{Hidden and coexistent magnetic phases in Kondo-type Cerium Hexaboride (CeB$_6$)}

\author{C. K. Barman}
\affiliation{Department of Physics, Indian Institute of Technology, Bombay, Powai, Mumbai 400 076, India}
\author{Prashant Singh}
\email{prashant@ameslab.gov}
\affiliation{Division of Materials Science $\&$ Engineering, Ames Laboratory, Ames, Iowa 50011, USA}
 \author{D. D. Johnson} 
 \email{DDJ@ameslab.gov}
\affiliation{Department of Materials Science $\&$ Engineering, Iowa State University, Ames, Iowa 50011, USA}
\affiliation{Division of Materials Science $\&$ Engineering, Ames Laboratory, Ames, Iowa 50011, USA}
\author{Aftab Alam}
\email{aftab@phy.iitb.ac.in}
\affiliation{Department of Physics, Indian Institute of Technology, Bombay, Powai, Mumbai 400 076, India}

\begin{abstract}
The heavy-fermion material CeB$_6$ shows hidden magnetic ordered phases. Besides Ferromagnetic (FM) and Antiferromagnetic (AFM) phases, CeB$_6$ is speculated to form a unique antiferroquadrupolar (AFQ) phase that is orbital in nature. Hidden from many characterization methods that cannot assess orbital ordering, debate continue on its origins. From electronic-structure calculations, we find that these phases are energetically almost degenerate, suggesting that magnetic domain walls form, possibly with defect boundaries. Only calculations with spin-orbit coupling reproduce most band structures and Fermi surfaces found in experiment, indicating crystal-field splitting is critical. Simulated ionization  (Ce f$^0$ and f$^1$) peaks also agree with photoemission. Small pressures stabilizes the AFM over FM, the observed phase at low temperature. Such small physical pressure may be realized from, e.g., intrinsic defects, such as vacancies, antisites, and surfaces.

\end{abstract} 
\date{\today}
\maketitle

\begin{figure*}[t]
\centering
\includegraphics[scale=0.30]{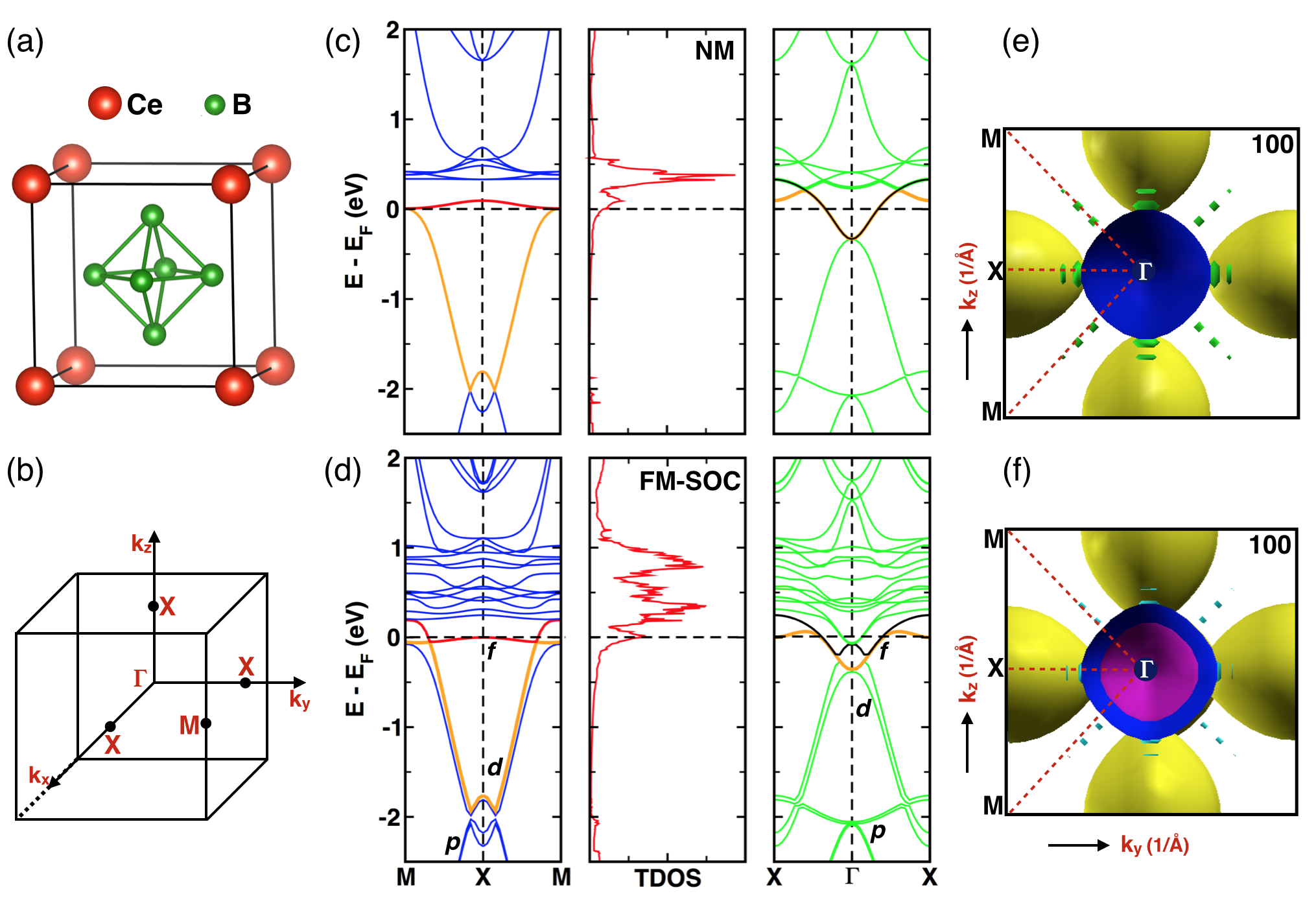}\
\caption {(Color online) For CeB$_{6}$, (a) crystal structure, (b) Brillouin zone (BZ) with high-symmetry points, and band-structures,  density of states (in states/eV-cell) and Fermi-surfaces for (c,e) non-magnetic and (d,f) magnetic spin-orbit cases.}
\label{Fig1}
\end{figure*}

{\par} Hidden-ordered phases are a well-known mystery of heavy fermion systems. Such phases remain hidden in most characterization techniques because they cannot assess orbital ordering. As such, the origin of such phases remains controversial. These peculiar phases are often found in rare-earth-based compounds, acquiring a rich, low-temperature (T) phase diagram; and it is speculated that 4$f$-electrons play a crucial role in their low-T structural and magnetic stability. CeB$_6$ is an archetypal example for magnetically hidden-ordered phases. It shows a unique antiferroquadrupolar (AFQ) ordering\cite{PS2009,YK2009} at temperature T${_Q}< 3.2$ K, associated with ordering of magnetic quadrupolar moments at cube corners with wave vector\cite{MSHI2001,HJ2014}  ${\bf K} \equiv [ \frac{1}{2}  \frac{1}{2} \frac{1}{2} ] $ in the cubic Brillouin zone (Fig.~\ref{Fig1}).  Quadrupolar ordering is orbital in nature, arising due to the distortion of electronic charge cloud of the unpaired electrons in their 4$f$ orbitals. Consequently, it is invisible to conventional neutron diffraction\cite{JME1985} and can only be visualized by resonant x-ray scattering or related probes that can tune to the orbital degrees of freedom.\cite{HN2001,TM2009,TM2012} In addition, conventional AFM order (with a double-${\bf K}$ commensurate  ${\bf K}_2=[ \frac{1}{4}  \frac{1}{4}  0 ]$ and ${\bf K}_{2}^{'}=[ \frac{1}{4}  \frac{1}{4}  \frac{1}{2} ]$ structure) is found below T$_N=2.3$ K.\cite{AS2013,OZ2003}
Neutron and Raman-scattering experiments\cite{EZ1984} show that multiplet J=5/2 in CeB$_6$ splits into a groundstate quartet ${\Gamma}_8$ and an excited doublet ${\Gamma}_7$ at 540 K in a cubic crystalline electric field. With the ${\Gamma}_8$ state having both magnetic and quadrupolar moments, the inter-site magnetic and quadrupolar interactions of the RKKY-type coexist.\cite{TSSS1998, SSS1999} The competition of these interactions results in a complex magnetic phase diagram involving FM, AFM and AFQ phases.\cite{JME1985,NS1984}
 

{\par} The ordering phenomena in CeB$_6$ is acknowledged to be governed by AFM interactions between multi-polar moments of the Ce-4$f$ electrons mediated by itinerant conduction electrons, which lift the degeneracy of the $\Gamma_8$ state of the Ce ions in their cubic crystal field.\cite{RS1997,PT2003}  Likewise, CeB$_6$ is known for its narrow electron spin resonance (ESR) signal,\cite{SV2006,SV2009} suggestive of FM correlations.\cite{PS2012} To understand its rich and complex magnetic low-T phases CeB$_{6}$ continues to attract great interest from experimentalists and theorists alike. Up to now, however, direct observation\cite{OZ2003} has remained elusive and theoretical models are lacking. The challenge is to describe coexistent, near-degenerate magnetic phases (FM, AFM and AFQ) of CeB$_6$ at low temperatures that arise from hybridization of Ce-4$f$ electrons with conduction electrons that lead to multi-polar ordering and the unusual magnetic phase diagram.

Recently, Jang, \emph{et al.}\cite{HJ2014} highlighted the FM correlations in CeB$_{6}$ and suggested an intimate interplay between the AFQ and AFM order parameters below T$_N$. Such dependence of ordering on FM correlations along with AFM coupling between dipolar and multi-polar moments of Ce-$4f$ electrons requires a more detailed understanding of the microscopic origin of the competing FM, AFM and AFQ  phases.   To fill this gap, we provide insight into the existence of these phases from the electronic structures.


{\par} {\it Computational method}: Calculations were carried out using the Vienna Ab-Initio Simulation Package (VASP) \cite{KH1993,KJ1999} based on density functional theory (DFT). We have employed generalized gradient approximation (GGA) by Perdew-Bueke-Ernzerhof (PBE),\cite{PBE1997} to describe the exchange correlation interactions. The valence interactions were described by projector augmented-wave (PAW) method \cite{PEB1994,KJ1999} adopting default kinetic energy cutoff (Ce 299.90 eV and B 318.606 eV) for the plane-wave orbitals. Total energies were converged up to $10^{-5}$ eV/cell. Spin-orbit coupling (SOC) was included self consistently within the scalar relativistic bands.

{\par}  CeB$_{6}$ crystallizes in a CsCl structure with Ce ions at cube corners and B$_6$ octahedra at its body center, see Fig.~\ref{Fig1}(a). The bulk Brillouin zone (BZ) is cubic, Fig.~\ref{Fig1}(b), with center, face-center, and edge-center denoted as $\Gamma$, $X$, and $M$ points, respectively. For CeB$_{6}$ we simulated nonmagnetic (NM), FM, AFM, AFQ and a new ordered phase, namely, Antiferro-tripolar (AFT). The AFT is nothing more than AFQ order with reduced magnetic periodicity at third neighboring Ce-atom rather than the fifth one (see supplement\cite{supl} for schematic). AFM, AFT and AFQ phases were simulated using $2\times 2 \times 2$, $3\times 3 \times 3$ and $5\times 5 \times 5$ supercell of the primitive CeB$_6$ ($7$-atom) unit cell, using the experimental lattice parameter $a=4.141~\AA$. \cite{LL2011}  We studied the stability and electronic structure of all  magnetic phases with(out) hydrostatic pressure. The BZ integrations were performed using Monkhorst-Pack meshes of $16^3$ for NM and FM, $4^3$ for AFM, $3^3$ for AFT and $2^3$ for AFQ.

\begin{table}[b]
\centering
\caption{CeB$_6$ energies (meV/atom) of AFM, AFT, AFQ and NM phases relative to FM phase at a$_{exp}= 4.141\AA$.\cite{LL2011} $SOC$ ($no-SOC$) are results with (without) spin-orbit coupling. }
\label {Table1}
\begin{tabular}{cccccccccc}
\hline
 &\hspace{0.3cm}FM \hspace{0.2cm} &\hspace{0.3cm}AFM \hspace{0.2cm} &\hspace{0.3cm}AFT \hspace{0.2cm} &\hspace{0.3cm}AFQ \hspace{0.2cm} &\hspace{0.3cm}NM \hspace{0.2cm}\\
\hline
no-SOC    & $0$ & $+1.6$ & $+1.3$ & $+1.5$ &$+3.7$ \\
\hline
 \, SOC       & $0$  & $+0.9$ & $+0.7$ & $+0.9$ & $-$   \\
\hline
\end{tabular}
\end{table}

{\par} {\it Results and Discussions}: Experimentally,\cite{PS2009,YK2009,HJ2014,OZ2003,AS2013}  the low-T magnetic phases of CeB$_6$ are the FM, AFM and AFQ that emerge in a very narrow low-T range ($< 3.2$ K).  However, there has always been discrepancies in the assessed relative stability of these phases. 
To address these issues, we calculated the relative stability of NM, FM, AFM, AFT and AFQ phases. Then, we investigated the electronic structure of the NM and lowest-energy state (FM) to shed some light on the experimental findings and the microscopic origin of the magnetism. 

{\par} Table~\ref{Table1} shows the energies of phases with(out) SOC relative to the FM phase (lowest-energy state). Notably, all magnetic phases are extremely close in energy -- within $2~me$V. With SOC, they are almost all energetically degenerate. This degeneracy  is the precursor of a magnetic phase instability in CeB$_6$ that infers the co-existence of FM, AFM, AFT and AFQ. This degeneracy also suggests the possibility of magnetic domain formation with selected regions involving different order, and the possibility of magnetic defect boundaries. We have also simulated the stability of various magnetic defect structures, partly motivated to find structures with  lower energy than the FM phase. However, all these defected structures (shown in Fig.~S2\cite{supl} by labels Def1, Def2, and Def3) are higher in energy, see supplement.\cite{supl} To assess localization effects of Ce-$4f$ electrons, we added PBE+U calculations\cite{PBE+U} for CeB$_6$, with a Hubbard U introduced in a screened Hartree-Fock manner. The calculations are done with three non-zero U values on Ce-$4f$ electrons, i.e., U $= 3, 4, 5$~eV, as also supported in Refs. \onlinecite{hubbardU}. The relative energetics (see supplement\cite{supl}) of the FM, AFM, AFT and AFQ magnetic phases remain close  (within a few meV) to that of the U=0 case, and hence our conclusion remains intact.

{\par}  Figure~ \ref{Fig1}(c) and \ref{Fig1}(d) shows the band structure and  density of states (DOS) for NM and FM-SOC states, respectively. The low-T phase and its electronic structure is mainly governed by the dispersive 5$d$ and flat 4$f$-bands, shown along M-X-M and X-$\Gamma$-X. The flat bands near the Fermi energy (E$_F$) and Fermi-surface (FS) arise purely from Ce-4$f$ states. In NM case, the weight of 4$f$ band lies above E$_F$. The dispersive $d$-band (at X-points) is found to be about ${-2.0}$ eV below E$_F$ and the dispersive B $2p$ bands  are located near the bottom of this $d$-band. These bands at/near X agree fairly well with experiment.\cite{NAZ2015} In contrast to previous calculations,\cite{NAZ2015} we find hole-like character near the FS at $\Gamma$. 
This shortcoming in the NM case arises mainly due to absence of the spin-orbit coupling which then misses the crystal-field splitting. Moreover the small, shallow and relatively heavy electron-like pocket found in ARPES measurement is also missing in the NM band structure. ARPES data\cite{NAZ2015} shows additional features comprising of strong momentum dependent enhancement of the quasi-particle density which are stronger near the $\Gamma$ and weak at X. 

{\par}Figure~\ref{Fig1}(e) and \ref{Fig1}(f) shows the NM and FM-SOC FS map for CeB$_6$. The calculated FS is in good agreement with the measurements.\cite{NAZ2015,AK2016} The FS indicates multiple hole pockets, including an oval shaped contour at X. The spectral intensities around $\Gamma$ are stronger compared to those of X. The two contours in SOC FS plot, blue and magenta around $\Gamma$ represents the band splitting. In Fig.~\ref{Fig1}(f), one can notice hole-like pocket at $\Gamma$, in contrast to NM case, with strongly renormalized bands corresponding to the observed, so-called, \emph{hot spots}.\cite{NAZ2015}

\begin{figure}[t]
\centering
\includegraphics[scale=0.30]{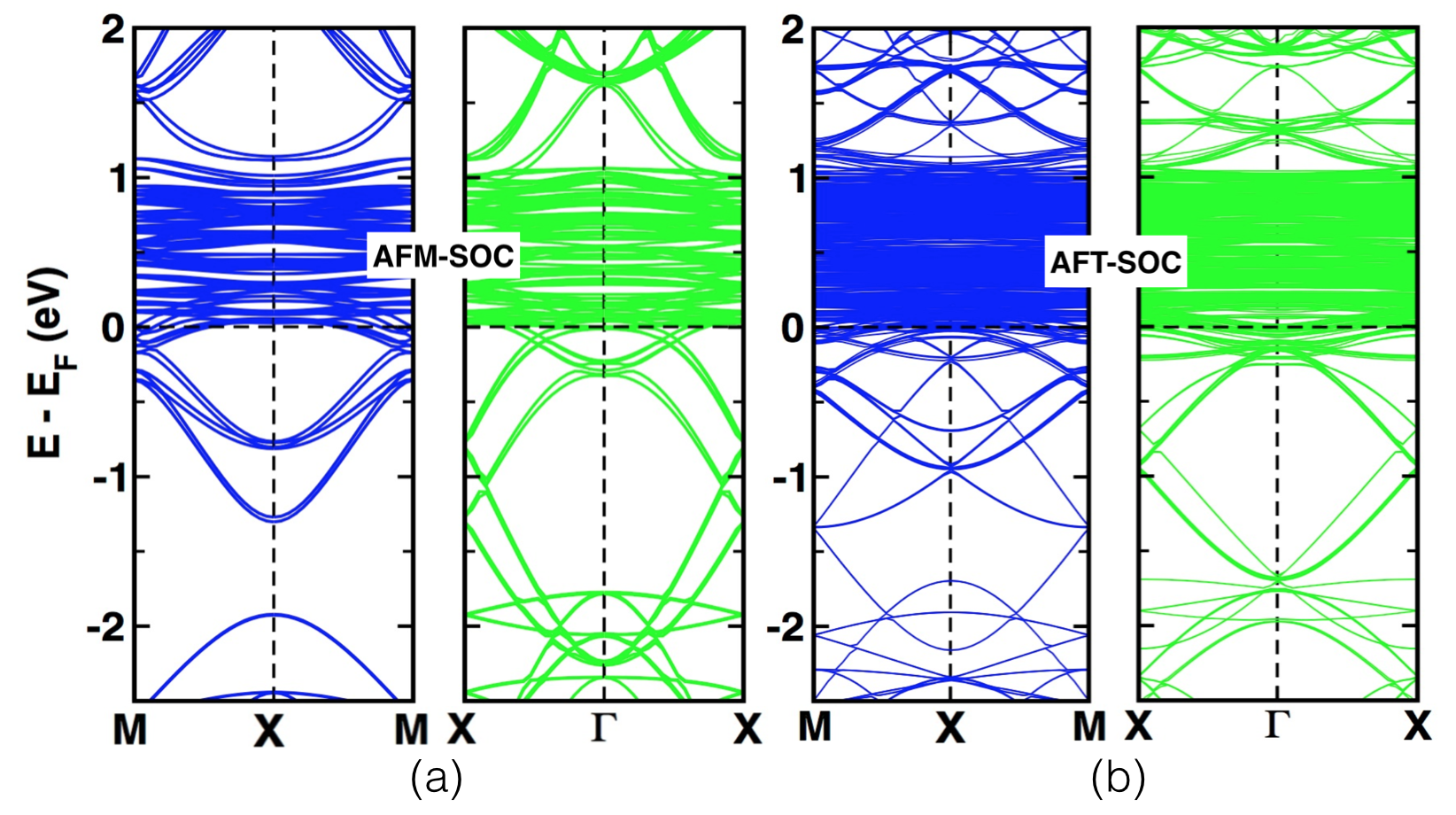}
\caption {Band-structure with spin-orbit coupling of (a) AFM, and (b) AFT phase of CeB$_{6}$ along M-X-M and X-$\Gamma$-X.} 
\label{Fig2}
\end{figure}

{\par} Including crystal-field splitting resolves most of the differences. Figure~\ref{Fig1}(d) shows the dispersion and DOS with SOC. One  immediately notices the location of flat Ce-4$f$ bands slightly below E$_F$, as observed from ARPES data, although their energy position differ slightly. The DOS shows similar behavior but with more electronic density below E$_F$. More importantly, parabolic shape band along X-$\Gamma$-X is found to form very close to E$_F$ at $\Gamma$ which emanates hole-like pocket, as observed, and corresponding to hot spots on the Fermi surface.\cite{NAZ2015} 

{\par}The origin of such hot spots is not clear; but, it is speculated that hot spots may arise due to the unusual low-T magnetic order observed in this system. In particular, hot spots could be related to the relatively high temperature FM fluctuation which is a precursor to other magnetic ordering emerging at lower temperature such as AFM, AFT or AFQ. The emergence of such low temperature magnetic order is highly possible because of their extremely close energetics compared to FM case, as shown in Table~\ref{Table1}. For completeness, we also calculate the band structure for AFM and AFT cases (Fig.~\ref{Fig2}). One can notice, strong renormalization of bands near E$_F$ at $\Gamma$-point in both these cases. There are several features in these bands which can be corroborated with those of ARPES data.\cite{NAZ2015} For example additional flat bands near $-1.8$ eV along X-$\Gamma$-X in AFT case. Parabolic shaped bands near E$_F$ at $\Gamma$-point which are relatively more flat  compared to those in ARPES data.\cite{NAZ2015} 

\begin{figure}[t]
\centering
\includegraphics[scale=0.22]{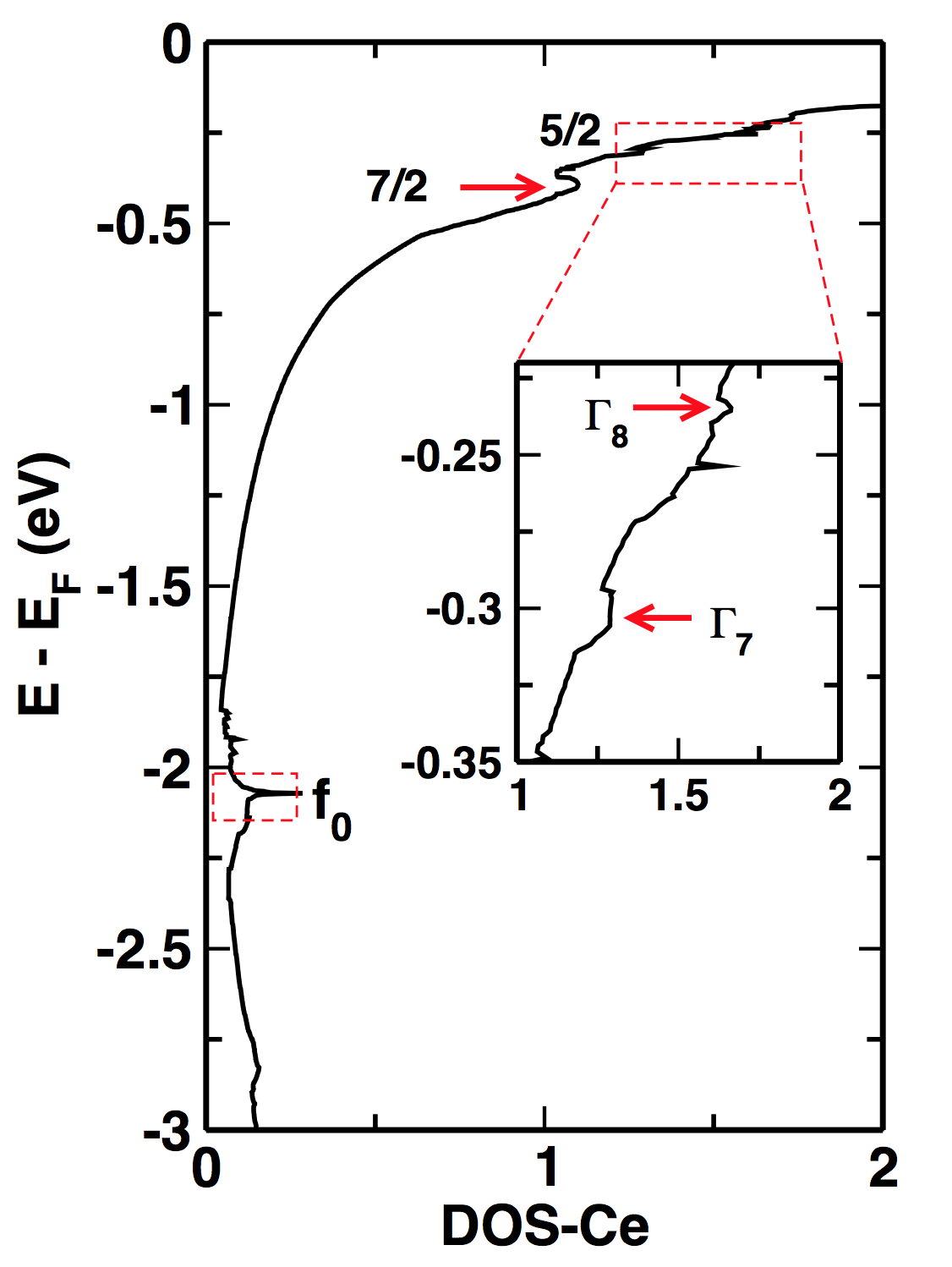}
\caption {Ce-projected density of state (DOS) in CeB$_{6}$. Non-dispersive flat Ce$-$4$f^{1}$ and broad Ce$-$4$f^{0}$ peaks agrees with experiments.\cite{NAZ2015} (Inset) J=5/2 level splitting into $\Gamma_{7}$ and  $\Gamma_{8}$.} 
\label{Fig3}
\end{figure}

\begin{table*}[t]
\centering
\caption{Relative energies (meV/atom) of AFM and AFT phases relative to FM phase under pressure projected via a volume change (or lattice parameter change $\Delta a/a_{exp}$). AFM stabilizes over FM upon $1\%$ reduction in $a$. Only z-component per Ce of the non-collinear spin $(\vec{\mu}_{spin}^{SOC})$ and orbital $(\vec{\mu}_{orbital}^{SOC})$  moments (Bohr magneton) are shown. Relatively small $x-$ and $y-$components are given in section IV of supplement\cite{supl}} 
\label {Table2}
\begin{tabular}{c|cccc|cccc|cccc|cccc|cccccccccccc}
    \hline
$\Delta{a}/a_{exp}$\hspace{0.3cm} &&$\Delta{E^{no-SOC}}$&&&&$\Delta{E^{SOC}}$&&&&$\mu_{spin}^{z(SOC)}$&&&&$\mu_{orbital}^{z(SOC)}$\\
  (in \%) \hspace{0.3cm}    & FM          &   AFM       &  AFT       &&   FM         &    AFM      &    AFT      &&  FM    &   AFM  &   AFT  &&  FM   &   AFM  &  AFT  \\
\hline                                                                            
$+2$\hspace{0.3cm}  &$0.0$ & $+70.14$  &  $+63.14$  &&$0.0$  &$+69.29$  &  $+62.57$  &&  0.78 &  0.74 &  0.77 && 0.80 &  0.94 &  1.00 \\
$+1$\hspace{0.3cm}   &  $0.0$   & $+38.57$ &  $+20.28$  &&  $0.0$  &$+37.57$  &  $+19.57$  &&  0.75 &  0.68 &  0.74 && 0.74 &  0.83 &  0.94 \\
~~~$0$\hspace{0.3cm}     &    $0.0$    & $+1.57$  &  $+1.29$  &&        $0.0$  & $+0.86$   &     $+0.71$  &&  0.71 &  0.59 &  0.70 && 0.68 &  0.71 &  0.85 \\
$-1$\hspace{0.3cm} & $0.0$& $-37.00$&   $+28.71$  &&  $0.0$  & $-37.57$  & $+28.14$  &&  0.67 &  0.49 &  0.63 && 0.61 &  0.57 &  0.73 \\
$-2$\hspace{0.3cm} & $0.0$ & $-70.71$ & $+135.0$ && $0.0$ &$-71.14$  & $+134.4$  &&  0.64 &  0.36 &  0.53 && 0.57 &  0.40 &  0.58 \\ 
    \hline
  \end{tabular}
\end{table*}

{\par} To locate the energy of Ce-4$f$ states, we present the SOC DOS of Ce (Fig.~\ref{Fig3}), showing a more localized f$^0$ ionization peak  at $-2.05~e$V, which overlaps with the bottom of the ellipsoid band. This position of f$^0$ peak agrees  with those of the integrated energy distribution from experiment.\cite{NAZ2015} Near E$_F$, there are screened f$^1$ states of Ce that splits into J$=5/2$ and $7/2$ components due to SOC. These states are located about $0.3-0.35$ eV below E$_F$. Interestingly, the $5/2$ state further splits into the crystal-field levels, i.e., the $\Gamma_7$ doublet and $\Gamma_8$ quartet (see inset). These two levels differ in energy by $62~me$V, agreeing fairly well with the $50~me$V from photoemission. \cite{NAZ2015,AK2016}

Pressure is a crucial factor that can change the electronic and/or magnetic structure of a material. We  studied the effect of pressure by varying the lattice constant $a$ from $-2\%$ to $+2\%$ relative to a$_{exp}$. Table~\ref{Table2} shows the energies relative to FM phase and spin and orbital moments of the phases, with and without spin-orbit coupling.  As discussed before, FM, AFM and AFT phases remain almost energetically degenerate (within a meV) at zero pressure (a$_{exp}$). Interestingly,  AFM is stabilized over FM  under a small hydrostatic pressure ($\Delta a$ change by $-1\%$ ). Such pressure may be realized under a variety of situations, e.g., (i) cell reduction by applied pressure; (ii) intrinsic defects, such as vacancies and antisites; and (iii) surface effects. Due to the large cell sizes of AFT and AFQ phases, inclusion of these effects are beyond the scope of the present study. However, stability of AFM over FM (or AFT) phase under such a small pressure do indicate the correct trend for the existence of AFM phase in the low-T range (T$<2.3 K$).\cite{OZ2003,AS2013}

In addition, CeB$_6$ is an intriguing heavy-fermion system in which the Ce orbital moment is comparable/larger in magnitude than its spin moment, as is obvious from Table~\ref{Table2}, where z-components of the non-collinear spin $(\vec{\mu}_{spin}^{SOC})$ and orbital $(\vec{\mu}_{orbital}^{SOC})$  moments reported.
 AFM state has a zero net moment in the cell, while other phases has a finite moment due to a lone Ce atom in the cell. With increasing $a$, Ce-4$f$ electrons become more localized leading to larger Ce-moment, whereas increasing pressure (reducing $a$) enhances hybridization between Ce-4$f$ and B-2$p$ orbitals which reduces Ce-moment, see Table~\ref{Table2}.

{\it Conclusion}: We have provided electronic insight to the debated origin of the competing low-T magnetic phases of the heavy-fermion hexaboride CeB$_{6}$ by detailing the electronic structure of the competing magnetic phases, including magnetic defect boundaries. The crystal-field splitting, controlled by spin-orbit coupling (SOC), yield electronic dispersion and Fermi surfaces (with correct electron and hole pockets) that agrees fairly well with those observed from ARPES, highlighting the importance of SOC in $f$-block systems and missed in previous calculations. Furthermore, our calculations reveal that dispersion around  $\Gamma$ in the BZ is strongly renormalized, as indicated by highly increased density of states there, which are observed as hot-spots in experiments. We also show that a small ($\le -1\%$ lattice contraction) applied hydrostatic pressure can lift the magnetic degeneracy of coexisting FM, AFM and AFQ phases. The change under pressure in the strength of hybridization between flat 4$f$-bands near the Fermi energy and low-lying dispersive 5$d$-bands was shown to play a crucial role in separating coexistent magnetic phase and stabilizing the AFM phase observed at low T. Moreover, keeping in mind the recent advent of topologically insulated phase in SmB$_{6}$, our study can open up a whole new search for topological insulator phase with magnetically active sites in CeB$_{6}$.

CKB acknowledges support from assistantship at IIT Bombay. Work at Ames Lab was supported by the U.S. Department of Energy (DOE),  Office of Science, Basic Energy Sciences, Materials Science and Engineering Division. Ames Laboratory is operated for the U.S. DOE by Iowa State University under Contract No. DE-AC02-07CH11358. 
\vspace{0.1cm}


\end{document}